\documentclass[a4paper]{jpconf}
\usepackage{graphicx}
\newcommand{\bra}[1]{\langle #1|}
\newcommand{\ket}[1]{|#1\rangle}
\newcommand{\braket}[2]{\langle #1|#2\rangle}

\def\Nj{\{N_j\}}
\def\Ntj{\{\tilde N_j\}}
\def\Qz{{\bf Q}}
\def\qj{{\bf q}_j}

\def\d{{\rm d}}

\def\e{{\rm e}}
\def\i{{\rm i}}
\def\Pro{{\sf P}}

\def\hpartj{\{h_{n_j}\}}
\def\ee{e$^+$e$^-$}
\def\ppb{${\rm p}\bar{\rm p}\;$}

\begin{document}
\title{What is the meaning of the statistical hadronization model?}

\author{Francesco Becattini}

\address{Universit\`a di Firenze and INFN Sezione di Firenze, Via G. Sansone
1, I-50019, Sesto F.no (Firenze), Italy}

\ead{becattini@fi.infn.it}

\begin{abstract}
The statistical model of hadronization succeeds in reproducing particle 
abundances and transverse momentum spectra in high energy collisions 
of elementary particles as well as of heavy ions. Despite its apparent 
success, the interpretation of these results is controversial and the 
validity of the approach very often questioned. In this paper, we would 
like to summarize the whole issue by first outlining a basic formulation 
of the model and then comment on the main criticisms and different kinds
of interpretations, with special emphasis on the so-called ``phase space
dominance". While the ultimate answer to the question why the statistical 
model works should certainly be pursued, we stress that it is a priority to 
confirm or disprove the fundamental scheme of the statistical model by 
performing some detailed tests on the rates of exclusive channels at lower 
energy.
\end{abstract}.

\section{Introduction}

The statistical model is a model of hadronization, thus aiming at reproducing 
the quantitative features of this process. Its founding ideas date back to 
Fermi \cite{fermi} and Hagedorn \cite{hage}, though a basic and precise formulation 
of this model has been lacking ever since; providing such a formulation is among 
the goals of the present work. 

In the statistical hadronization model (SHM), the physical picture of a high 
energy collision is that of a QCD-driven dynamical process eventually giving rise 
to the formation of extended massive objects (called {\em clusters} or {\em fireballs}) 
which decay into hadrons in a purely statistical fashion. The number, as well as 
the kinematical and internal quantum properties of these objects are determined by 
the previous dynamical process and are thus not predictable within the SHM itself; 
they can be hopefully calculated with perturbative QCD, like in other hadronization 
cluster models \cite{herwig}. A distinctive feature of the statistical model 
in comparison with other cluster models is that clusters have a finite spacial 
extension. This is actually a crucial assumption, the one which ultimately allows to 
make calculations. 

Probably, the best known model with relativistic extended massive objects
is the bag model \cite{bag} and indeed the SHM can be considered as a model for 
the strong decays of bags. On the other hand, on the experimental side, there is 
now strong evidence of the finite extension of the hadron emitting sources in high 
energy collisions, from the observed quantum interference effects in the production 
of identical particles. It is therefore reasonable to take a finite volume of the
hadron sources as a key ingredient for a hadronization model.

In this paper, we first expound a precise formulation of the model starting from
the very basic assumption of local statistical equilibrium (Sect.~2). We then
comment on various criticisms and interpretations of the successes of this model, 
especially in elementary collisions (Sect.~3). The paper is concluded with 
a discussion about the possible fundamental physical meaning of the model 
(Sect.~4).

\section{The statistical model: a fundamentalist approach}

The basic idea of the model is very simple and it lies in two assumptions. 
The first is that in the late stage of a high energy collision, some extended 
massive objects, defined as {\em clusters}, are produced which decay into hadrons 
at a critical value of energy density or some other relevant parameter. The 
second, fundamental,
assumption of the statistical model is that {\em all multihadronic states within 
the cluster compatible with its quantum numbers are equally likely}. This makes 
the model predictive as to the production rates of hadrons and resonances from 
the clusters so that SHM might be regarded as an effective model of the decays of 
hadronic extended relativistic massive objects. This idea was essentially introduced 
within the statistical bootstrap model by Hagedorn \cite{hage}, who, by identifying 
clusters with massive resonances, predicted the hadronic mass spectrum to rise 
exponentially. This seems to be still a very succesful prediction \cite{bronio}, 
but is neither implied nor required by the SHM alone: in principle, clusters need
not to be identified with actual resonances to decay statistically.    

Despite the apparent simplicity of the key assumption of the SHM, it is not
as straightforward as it might seem at first sight to calculate the cluster decay
rates into different multi-hadronic states. These difficulties arise from the fact 
that the basic postulate only tells us that localized states are equiprobable, yet 
these states are essentially different from the observable asymptotic states. 
As we will show, such difference is not an issue when the volume is sufficiently large 
and can be disregarded in most applications where the canonical or grand-canonical 
ensemble are used, but it is relevant at a fundamental level of description and 
must be taken into account when the volume is small, i.e. less than ${\cal O}(10)$ 
fm$^3$.

Suppose that we can describe the cluster as a mixture of localized multi-hadronic
states $\ket{h_V}$ and, according to the basic assumption, all states have the same 
statistical weight. Then, one can write down a {\em microcanonical partition 
function} as:
\begin{equation}\label{mpf}
 \Omega = \sum_{h_V} \bra{h_V} \Pro_i \ket{h_V}
\end{equation}
where $\Pro_i$ is the projector over all conserved quantities in strong interactions, 
namely energy-momentum, angular momentum, parity, isospin etc. 
It must be emphasized that these states $\ket{h_V}$ are {\rm not} the asymptotic 
observable free states of the Fock space which can be labelled with particle 
multiplicities for each species $\{N_1,N_2,\ldots,N_K\} \equiv \Nj$, their momenta 
and helicities. Thus, the probability of observing a set of particles with four-momenta 
$p_1,\ldots,p_N$ is {\em not} $\bra{h_V} \Pro_i \ket{h_V}$ and, moreover, it cannot 
be obtained unambiguously from the Eq.~(\ref{mpf}). 

In order to define a suitable probability of observing an asymptotic multi-hadronic
state $\ket{f}$, one can recast the microcanonical partition function (\ref{mpf}) by 
using the completeness of states $\ket{f}$'s:
\begin{eqnarray}\label{mpf2}
 \Omega &=& \sum_{h_V} \bra{h_V} \sum_f \ket{f} \bra{f} \Pro_i \ket{h_V} \nonumber \\
        &=& \sum_f \bra{f} \Pro_i \sum_{h_V} \ket{h_V} \braket{h_V}{f} \equiv 
	\sum_f \bra{f} \Pro_i \Pro_V \ket{f}
\end{eqnarray}
where $\Pro_V = \sum_{h_V} \ket{h_V} \bra{h_V}$ is the projector on the localized
states. We note that the last expression of 
$\Omega$ in Eq.~(\ref{mpf2}) is a proper trace, whereas it was not in Eq.~(\ref{mpf}) 
as the states $\ket{h_V}$ do not form a complete set, i.e. they are not a basis of the 
Hilbert space. Looking at Eq.~(\ref{mpf2}), it is tempting to set the probability 
$\rho_f$ of the state $\ket{f}$ as proportional to $\bra{f} \Pro_i \Pro_V \ket{f}$. 
Yet, one could have worked out $\Omega$ differently from Eq.~(\ref{mpf2}), for 
instance (if $\ket{h_V}$ are properly normalized):
\begin{eqnarray}\label{mpf3}
 \Omega &=& \sum_{h_V} \bra{h_V} \Pro_i \ket{h_V} =  
 \sum_{h_V} \bra{h_V} \Pro_V \Pro_i \ket{h_V} \nonumber \\
  &=& \sum_{h_V} \bra{h_V} \sum_f \ket{f} \bra{f} \Pro_V \Pro_i \ket{h_V} \nonumber \\ 
  &=& \sum_{f} \bra{f} \Pro_V \Pro_i \sum_{h_V} \ket{h_V} \braket{h_V}{f} =  
      \sum_{f} \bra{f} \Pro_V \Pro_i \Pro_V \ket{f}
\end{eqnarray}
whence $\rho_f$ could now be set as proportional to $\bra{f} \Pro_V \Pro_i \Pro_V 
\ket{f}$. The latter expression is different from $\bra{f} \Pro_i \Pro_V \ket{f}$ 
unlike $[\Pro_i,\Pro_V] = 0$, which is not the case as long as $\Pro_i$ includes conservation 
of energy and momentum (we will see this more in detail in the next section). 

Which of the two probabilities:
\begin{eqnarray}\label{probs}
  \bra{f} \Pro_V \Pro_i \Pro_V \ket{f} \qquad  {\rm and} \qquad 
  \bra{f} \Pro_i \Pro_V \ket{f} 
\end{eqnarray}
is the correct one? A well-defined probability should meet two requirements:
\begin{itemize}
\item{} positivity;
\item{} respect conservation laws, i.e. $\rho_f = 0$ if $\ket{f}$ has not the same
quantum numbers as the initial state; in other words, $\rho_f=0$ if $\Pro_i \ket{f} =0$.
\end{itemize}
The leftmost expression in Eq.~(\ref{probs}) fulfills the positivity requirement
in that:
\begin{equation}
 \bra{f} \Pro_V \Pro_i \Pro_V \ket{f} = \bra{ \Pro_V f} \Pro_i \ket{\Pro_V f} = 
 a^{-1} \bra{\Pro_V f} \Pro^2_i \ket{\Pro_V f} = a^{-1} \braket{\Pro_i \Pro_V f}
 {\Pro_i \Pro_V f} \ge 0
\end{equation}
where we have used the hermiticity of $\Pro_i, \Pro_V$ and the fact that $\Pro^2_i = 
a \Pro_i$ through a positive divergent constant $a$ \footnote{The divergence of this
constant is owing to the non-compactness of the Poincar\'e group. This can be
understood by considering the projector on energy-momentum $\delta^4 (P - P_{\rm op})$}.
However, the second requirement is not fulfilled: $\rho_f$ turns out to be not vanishing
even for states $\ket{f}$ which do not have the same energy-momentum as the initial
state.

On the other hand, the rightmost expression in (\ref{probs}) manifestly fulfills 
the conservation requirement because $\Pro_i$ has $\ket{f}$ as its argument,
but, on the other hand, it is not positive definite. Positivity is recovered by 
changing the rightmost expression in Eq.~(\ref{probs}) into:
\begin{equation}\label{best}
  \rho_f \propto \bra{f} \Pro_i \Pro_V \Pro_i \ket{f} 
\end{equation}
i.e. by plugging one more projection operator on the initial state. Thereby:
\begin{equation}
 \bra{f} \Pro_i \Pro_V \Pro_i \ket{f} = \bra{\Pro_i f} \Pro_V \ket{\Pro_i f} =  
 \bra{\Pro_i f} \Pro^2_V \ket{\Pro_i f} = \braket{\Pro_V \Pro_i f}{\Pro_V \Pro_i f} 
 \ge 0
\end{equation}
where now we have used the idempotency of $\Pro_V$. The definition (\ref{best}) of
the probability leads to a microcanonical partition function which differs from
the proper one (\ref{mpf}) just by a positive (divergent) constant which is anyhow 
irrelevant for the calculation of averages. Indeed:
\begin{equation}
 \tr (\Pro_i \Pro_V \Pro_i) = \tr (\Pro^2_i \Pro_V) = a \tr (\Pro_i \Pro_V) = 
 a \Omega
\end{equation}
where Eq.~(\ref{mpf2}) has been used.    

The leftmost definition of the probability in Eq.~(\ref{probs}) was in fact used 
in ref.~\cite{bf1} to work out the rates of multi-hadronic exclusive channels 
$\Nj$ in the SHM. Although breaking energy-momentum conservation, the expressions 
of these rates are the same as those obtained from integrating Eq.~(\ref{best}) 
over momenta of final particles. An attractive feature of this definition is that 
it can be written {\em formally}, with $\Pro_i \equiv \ket{i}\bra{i}$, as:
\begin{equation}
 \rho_f \propto \left| \bra{f} \Pro_V \ket{i} \right|^2
\end{equation}
i.e. it looks similar to a transition probability. However, the expression (\ref{best}) 
seems to be the best suited because it naturally meets both aforementioned basic 
requirements. 

According to the definition (\ref{best}), the cluster is regarded as the mixture 
of states:
\begin{equation}\label{mix1}
  \sum_{h_V} \Pro_i \ket{h_V} \bra{h_V} \Pro_i
\end{equation}
unlike in the leftmost definition in Eq.~(\ref{probs}), where the mixture
turns out to be:
\begin{equation}\label{mix2}
   \sum_{h_V} \ket{h_V} \bra{h_V} \Pro_i \ket{h_V} \bra{h_V}
\end{equation}
We think that a mixed state where $\Pro_i \ket{h_V}$ are equiprobable, like in
(\ref{mix1}) is the most appropriate definition of microcanonical ensemble because 
any state in the mixture has actually the same quantum numbers, including energy 
and momentum, of the cluster.

\subsection{The microcanonical ensemble}

Once a satisfactory definition of the probability of observing an asymptotic 
multi-hadronic state is found, that is Eq.~(\ref{best}), we can start working it
out to obtain tractable formulae. We start by first developing the projector 
$\Pro_i$ defining the microcanonical ensemble. 

In principle, in the microcanonical ensemble, all conserved quantities should be 
included: energy, momentum, angular momentum, parity, internal charges and 
C-parity (if the cluster is neutral). The correct way to implement these conservation
laws is to project the multi-particle states onto the irreducible state of the 
full symmetry group which defines the initial state, i.e. the hadronizing cluster.
The full symmetry group is the product of the extendend Poincar\'e group 
IO(1,3)$^\uparrow$, the isospin SU(2), the U(1)'s related to conserved additive
charges and the discrete group Z$_2$ of charge conjugation if the initial state 
is neutral. Accordingly, the projector $\Pro_i$ can be factorized as:
\begin{equation}\label{progen}
   {\sf P}_i = {\sf P}_{P,J,\lambda,\pi} {\sf P}_\chi {\sf P}_{I,I_3} {\sf P}_\Qz  
\end{equation}
where $P$ is the four-momentum of the cluster, $J$ its spin, $\lambda$ its 
helicity, $\pi$ its parity, $\chi$ its C-parity, $I$ and $I_3$ its isospin and 
its third component and $\Qz = (Q_1,\ldots,Q_M)$ a set of $M$ abelian (i.e.
additive) charges such as baryon number, strangeness, electric charge etc.
Of course, the projection ${\sf P}_\chi$ makes sense only if $I = 0$ and
$\Qz = {\bf 0}$; in this case, ${\sf P}_\chi$ commutes with all other projectors. 
 
The projector ${\sf P}_{P,J,\lambda,\pi}$ onto the irreducible state (transforming
according to an irreducible unitary representation $\nu$ of IO(1,3)$^\uparrow$) 
with definite four-momentum, spin, helicity and parity can be written by using 
the normalized invariant measure $\mu$ of the Poncar\'e group as:
\begin{equation}\label{proj}
 \Pro_{P,J,\lambda,\pi} = \frac{1}{2} \sum_{z={\sf I,\Pi}} \dim \nu 
 \int \d \mu(g_z) \; D^{\nu \dag}(g_z)^i_i \, U(g_z)   
\end{equation} 
where $z$ is the identity or space inversion ${\sf \Pi}$, $g_z \in 
{\rm IO}(1,3)^\uparrow_{\pm}$, $D^\nu(g_z)$ is the matrix of the irreducible 
representation $\nu$ the initial state $i$ belongs to, and $U(g_z)$ is the unitary 
representation of $g_z$ in the Hilbert space. Similar integral expressions can be 
written for the projectors onto internal charges, for the groups SU(2) (isospin) 
and U(1) (for additive charges). Although projection operators cannot be rigorously
defined for non-compact groups, such as Poincar\'e group, we will maintain this naming 
relaxing mathematical rigour. In fact, for non compact-groups, the projection 
operators cannot be properly normalized so as to ${\sf P}^2 = {\sf P}$ and this 
is indeed related to the fact that $| i \rangle $ has infinite norm. Still, 
we will not be concerned with such drawbacks thereafter, whilst it will be favourable 
to keep the projector formalism. Working in the rest frame of the cluster, with 
$P = (M, {\bf 0})$, the matrix element $D^{\nu\dag}(g_z)^i_i$ vanishes unless the 
Lorentz transformations are pure rotations and this implies the reduction of the 
integration in (\ref{proj}) from IO(1,3)$^\uparrow$ to the subgroup ${\rm T}(4) 
\otimes {\rm SU}(2) \otimes {\rm Z}_2$ \cite{bf1}. In fact, the general transformation
of the extended Poincar\'e group $g_z$ may be factorized as:
\begin{equation}
  g_z = {\sf T}(x) {\sf Z} {\sf \Lambda} = {\sf T}(x) {\sf Z} 
  {\sf L}_{\hat{\bf n}}(\xi) {\sf R}
\end{equation}
where ${\sf T}(x)$ is a translation by the four-vector $x$, ${\sf Z} = {\sf I, \Pi}$
is either the identity or the space inversion and ${\sf \Lambda} = 
{\sf L}_{\hat{\bf n}}(\xi) {\sf R}$ is a general orthocronous Lorentz transformation 
written as the product of a boost of hyperbolic angle $\xi$ along the space-like 
axis $\hat{\bf n}$ and a rotation $\sf R$ depending on three Euler angles. Thus
Eq (\ref{proj}) becomes:
\begin{eqnarray}
 {\sf P}_{P,J,\lambda,\pi} &=& \frac{1}{2} \sum_{{\sf Z}={\sf I, \Pi}} 
 \frac{\dim \nu}{(2\pi)^4} \int \d^4 x \int \d {\sf \Lambda} \; D^\nu({\sf T}(x)
 {\sf Z}{\sf \Lambda})^{i*}_i \, U({\sf T}(x) {\sf Z} {\sf \Lambda}) \nonumber \\
 &=& \frac{1}{2} \sum_{{\sf Z}={\sf I, \Pi}} 
  \frac{\dim \nu}{(2\pi)^4} \int \d^4 x \int \d {\sf \Lambda} \; 
 \e^{\i P \cdot x} \pi^{z} D^\nu({\sf \Lambda})^{i*}_i \; U({\sf T}(x)) 
 \, U({\sf Z}) \, U({\sf \Lambda}) 
\end{eqnarray} 
where $z=0$ if ${\sf Z}= {\sf I}$ and $z=1$ if ${\sf Z}= {\sf \Pi}$.
In the above equation, the invariant measure $\d^4 x$ of the translation subgroup 
has been normalized with a coefficient $1/(2\pi)^4$ in order to yield a Dirac delta, 
as shown later. Furthermore, $\d {\sf \Lambda}$ is meant to be the invariant normalized 
measure of the Lorentz group, which can be written as \cite{lgroup}:
\begin{equation}\label{measure}
  \d {\sf \Lambda} = \d {\sf L}_{\bf n}(\xi) \, \d {\sf R} =
  \sinh^2 \xi \d \xi \, \frac{\d \Omega_{\hat{\bf n}}}{4 \pi} \, \d {\sf R}
\end{equation}
$\d {\sf R}$ being the well known invariant measure of SU(2) group, $\xi \in [0,+\infty)$
and $\Omega_{\hat{\bf n}}$ are the angular coordinates of the vector ${\bf n}$. 

If the initial state $| i \rangle $ has vanishing momentum, i.e. $P=(M,{\bf 0})$, 
then the Lorentz transformation $\sf \Lambda$ must not involve any non-trivial boost 
transformation with $\xi \ne 0$ for the matrix element $D^\nu({\sf \Lambda})^{i*}_i$ not to 
vanish. Therefore $\sf \Lambda$ reduces to the rotation ${\sf R}$ and we can write:
\begin{equation}
 {\sf P}_{P,J,\lambda,\pi} = \frac{1}{2} \sum_{{\sf Z}={\sf I, \Pi}} \frac{1}{(2\pi)^4} 
 \int \d^4 x \; (2J + 1) \int \d {\sf R} \; \e^{\i P \cdot x} \pi^{z}  
 D^J({\sf R})^{\lambda *}_\lambda \, U({\sf T}(x)) \, U({\sf Z}) \, U({\sf R}) 
\end{equation} 
Since $[{\sf Z}, {\sf R}] = 0$, we can move the $U({\sf Z})$ operator to the right of
$U({\sf R})$ and recast the above equation as:
\begin{eqnarray}\label{final}
 {\sf P}_{P,J,\lambda,\pi} &=& \frac{1}{(2\pi)^4} 
 \int \d^4 x \; \e^{\i P \cdot x} U({\sf T}(x)) (2J + 1) \int \d{\sf R} \; 
 D^J({\sf R})^{\lambda *}_\lambda \, U({\sf R}) \, 
 \frac{{\sf I} + \pi U({\sf \Pi})}{2}  \nonumber \\
   &=& \delta^4( P - P_{\rm op}) (2J + 1) \int \d{\sf R} \; 
 D^J({\sf R})^{\lambda *}_\lambda \, U({\sf R}) \, 
 \frac{{\sf I} + \pi U({\sf \Pi})}{2}
\end{eqnarray} 
The Eq.~(\ref{final}) is indeed the final general expression of the projector defining 
the proper microcanonical ensemble with $P=(M,{\bf 0})$, in which all conservation laws 
related to space-time symmetries are taken into account. The appeal of the above 
expressionresides in the factorization of projection operators onto the energy-momentum $P$, 
spin-helicity $J,\lambda$ and parity $\pi$ of the cluster. 

Also the projectors onto isospin $\Pro_{I,I_3}$ and onto additive charges $\Pro_{\Qz}$
in Eq.~(\ref{progen}) can be given an integral expression by using the invariant SU(2)
and U(1) group measures. The projector on a state with definite C-parity $\chi$ can be 
simply written as $({\sf I}+\chi {\sf C})/2$ where ${\sf C}$ is the charge-conjugation 
operator.

In most calculations, conservation of angular momentum, isospin, parity and C-parity 
is disregarded and only energy-momentum and abelian charges conservation is enforced. 
\footnote{Note that the relevant set of states is still defined microcanonical 
ensemble and we will comply with this tradition.} This is expected to be an 
appropriate approximation in high energy collisions, where many clusters are formed 
and the neglected constraints should not play a significant role. 
On the other hand, they are important in very small hadronizing systems (e.g. \ppb at 
rest \cite{heinz}) whereby the full projection operation in Eq.~(\ref{progen}) should 
implemented. For the restricted microcanonical ensemble, it can be easily seen from 
Eqs.~(\ref{progen},\ref{final}) that the projector can be rewritten as:
\begin{equation}\label{delta}
  \Pro_i = \delta^4(P - P_{\rm op}) \, \delta_{\Qz,\Qz_{\rm op}}
\end{equation}
and, if $\ket{f}$ is an eigenstate of four-momentum and charges, Eq.~(\ref{best}) as:
\begin{equation}\label{best2}
  \rho_f \propto \omega \, \delta^4(P - P_f) \, \delta_{\Qz,\Qz_f} 
  \, \bra{f} P_V \ket{f}
\end{equation}
where $\omega$ is a divergent constant, i.e. the whole space-time volume. 

\subsection{The effects of finite volume}

The second part of the calculation involves the projection onto localized states.
The projector $\Pro_V$ can be written as:
\begin{equation}\label{pv1}
      \Pro_V = \sum_{\Ntj,k} \ket{\Ntj,k}\bra{\Ntj,k}
\end{equation}
where $\Ntj$ are the occupation numbers of the particles {\em in the cluster} and
$k$ the variables labelling their kinematical modes, e.g. three integers
in case of a parallelepipedon box with fixed or periodic boundary conditions.
It should be pointed here that in the usual statistical model calculations, the 
interactions are taken into account by including all known resonances as free 
particles (in the cluster) with a distributed mass, according to a formalism 
developed by Dashen, Ma and Bernstein \cite{bdm, hagelect}.
In principle, the use of (\ref{pv1}) to calculate probabilities like in (\ref{best}) 
entails some difficulty because a localized state $\ket{\Ntj,k}$ is {\em not} an 
eigenstate of the actual particle number, which is defined in terms of the
operators creating and destroying free asymptotic states over the whole space.
In fact, a $N$-pion state in the cluster has non vanishing components on {\em all} 
free states of the pion field, i.e. on the states with $0, 1, 2, \ldots$ pions. 
Therefore, the projection should be performed in a full quantum relativistic field 
approach by identifying localized states as states of the quantum fields associated 
to particles and vanishing out of the cluster region. Hence, the projector $\Pro_V$ 
should be rather written as, in case of only one scalar particle:
\begin{equation}\label{pv2}
   \Pro_V = \int_V D \psi \ket{\psi}\bra{\psi}
\end{equation}
and Eq.~(\ref{best}) developed accordingly. In Eq.~(\ref{pv2}) $\ket{\psi} \equiv 
\otimes_{{\bf x}} \ket{\psi({\bf x})}$ and $D \psi$ is the functional measure; 
the functional integration must be performed over all functions having as support 
the cluster region $V$. Altogether, determining production rates involves the
calculation of the statistical mechanics of a field in the microcanonical ensemble.

Nevertheless, if the cluster size is sufficiently larger than the Compton wavelenght
of the particles involved, quantum field corrections are expected to be small
and the eigenstates of particle number operators in the whole space essentially 
correspond to those of the particle number operators in the cluster. Otherwise
stated, even though a localized $N$-pion state has non vanishing components 
on all free states of the pion field, the dominant one will be that on the asymptotic
$N$ pion states. The largest implied Compton wavelenght in a multi-hadronic system
is indeed the pion's one $\lambda_\pi \simeq 1.4$ fm; this is the minimal size of 
the cluster below which quantum field corrections cannot be disregarded. 

Hence, for clusters which are sufficiently larger than $\lambda_\pi$ we can use the 
approximation:
\begin{equation}\label{appr}
  \braket{\Nj,p}{\Ntj,k} \ne 0 \qquad {\rm iff} \;\; \Nj=\Ntj  
\end{equation}
where $p$ labels the set of kimatical variables (namely momenta and helicities) 
for the asymptotic free states $\ket{f} = \ket{\Nj,p}$. Now, by using Eq.~(\ref{appr})
the probability of a final state (\ref{best2}) can be calculated and reads:
\begin{equation}    
  \rho_{\Nj,p} \propto \delta^4(P - \sum_i p_i ) \, \delta_{\Qz_i,\sum_j \qj} 
  \sum_k \big| \braket{\Nj,p}{\Nj,k} \big|^2
\end{equation}
where the $p_i$'s are the four-momenta of the particles in the final state.
The rightmost factor in the above equation can be calculated as a cluster decomposition
and, in the framework of non-relativistic quantum mechanics, a relevant expression has 
been obtained in ref.~\cite{hagechai} in the limit of large volumes and in 
ref.~\cite{bf1} taking into account the finite volume. If $j$ labels the hadron
species $j=1,\ldots,K$ and $p$ now the set of particles' four-momenta:
\begin{equation}\label{clust}
 \rho_{\Nj,p} \propto\ \delta^4 (P-\sum_i p_i) \prod_j \sum_{\hpartj} 
 \frac{(\mp 1)^{N_j + H_j} (2J_j + 1)^{H_j}}
 {\prod_{n_j=1}^{N_j} n_j^{h_{n_j}} h_{n_j}!} \prod_{l_j=1}^{H_j} F_{n_{l_j}}
\end{equation}
where $\hpartj$ is a partition of the integer $N_j$ in the multiplicity representation
, i.e. $N_j = \sum_{n_j=1}^{N_j} n_j h_{n_j}$; $H_j = \sum_{n_j=1}^{N_j} h_{n_j}$ and:
\begin{equation}\label{fint}
 F_{n_l} = \prod_{i_l=1}^{n_l} \frac{1}{(2\pi)^3} \int_V \d^3 {\rm x} \; 
 \e^{\i {\bf x \cdot}({\bf p}_{c_l(i_l)}-{\bf p}_{i_l})}
\end{equation}
are integrals over the cluster region $V$, $c_l$ being the cyclic permutation
of the integers $1,\ldots,n_l$. For large volumes, the dominant term in the 
cluster decomposition (\ref{clust}) is obtained by taking $\hpartj = (N_j,0,\ldots,0)$,
implying $c_l \equiv I$ and reads:
\begin{equation}\label{phspa0} 
 \rho_{\Nj,p} \propto \Bigg[ \prod_j \frac{V^{N_j}(2J_j + 1)^{N_j}}{(2\pi)^{3N_j} N_j!}
 \Bigg] \delta^4(P - \sum_i p_i) 
\end{equation}
%

\section{Phase space dominance, Lagrange multipliers and all that}

Despite its apparent success in reproducing observables related to hadronization
process like particle multiplicities and transverse momentum spectra \cite{beca,becah,
becagp,becabiele} the statistical model is not very popular among high energy 
physicists. Besides the fact that, thus far, the featured apparent statistical 
equilibrium is not derivable from QCD (and it will remain so for a probably long 
time), one of the most bothering point seems to be the presence of thermodynamical 
quantities and chiefly temperature. 
Moreover, this fairly constant hadronization temperature (i.e. around 160 MeV 
\cite{becabiele}) is amazingly close to the estimated critical temperature of QCD 
and this obviously raises the question of its meaning.  

In the basic microcanonical formulation of the model, described in 
the previous sections, one deals with mass and volume of clusters, but it can be 
easily shown that if those become sufficiently large one can perform calculations 
in the canonical ensemble, which is far easier to handle, thereby introducing 
temperature through a saddle-point expansion \cite{bf1,bf2}. 
In the case of hadron gas, this is possible at relatively low values of masses and 
volumes \cite{bf2,liu}, around 8 GeV and 20 fm$^3$. Yet, talking about temperature 
in such small systems seems to be daring for the received wisdom of most physicists 
who, as soon as the word temperature is spoken, are led to think of a large system 
which has undergone a long cooking process before reaching equilibrium. It is widely
believed (the author is included) that this cannot occur after hadronization at a 
level of formed hadrons through inelastic collisions: the system expands too quickly
to allow this. If statistical equilibrium is genuine, it must be an inherent property 
of hadronization itself, i.e. hadrons are born at equilibrium as stated by Hagedorn 
many years ago \cite{hagelect} and reaffirmed by others more recently \cite{becah,
heinz2,stock,satz}.

Therefore, there have been some attempts to account for the success of the 
statistical model whose conclusions may be roughly clustered as follows: 
\begin{enumerate}
\item{} the results of the statistical model can be obtained from other models with
some supplementary assumption or invoking some special, so far neglected, mechanism;  
\item{} the statistical model grasps some truth of the hadronization process, 
but the apparent thermal-like features are an effect of a special property of the
quantum dynamics governing hadronization, which tends to evenly populates all final 
states: this is defined as {\em phase space dominance};   
\item{} the results of the statistical model are somehow trivial, due to the large 
multiplicities involved which eventually make the multi-hadronic phase space almost 
evenly populated.
\end{enumerate}

In the following I will comment on specific papers discussing this subject, whose 
attitude, on the basis of my personal understanding, is assigned to one (or more) of 
the previous points. I apologize in advance with the quoted authors for possible 
misunderstanding and too limited summary of their thought.

Ideas of the class (i) are proposed e.g. in refs.~\cite{bial,flork}. The typical 
exponential shape of the thermal spectra are explained in the framework of the 
string model, by adding to the basic picture additional fluctuations of the string 
tension parameter $\kappa$. The effect of the fluctuations is to broaden the 
gaussian shape 
of the $p_T$ spectra in the string model, turning it into an exponential one. 
Of course, this mechanism is one of the possible choices of nature, though very 
difficult to disprove. In general, it is certainly possible to make an existing
model more complicated to account for some otherwise more straightforward result 
in another model, and this is precisely where the criterium called Occam razor 
intervenes: between two models equally able to explain observations, the most 
economical should prevail. It is fair to say here that the string model has been 
tested against more observables and that the SHM should be tested against the 
same set of observables. Still, it is also true that the effective implementations 
of the string models are plagued by the need of many free parameters to reproduce 
the data and this raises many doubts about its predictive power \cite{delphi}. 
In fact, there is an ongoing work \cite{bf2,gabb} to implement SHM as hadronization 
model in an event generator to allow testing observables other than multiplicities 
and single particle inclusive transverse momentum spectra.

The paper by Hormuzdiar {\it et al}~\cite{oregon} is the one where the idea (ii) is
certainly argued more in detail. The basis of the whole argument is the similarity 
between the (classical) phase space of the set of particles $\Nj$, obtained by 
integrating (\ref{phspa0}) over momenta:
\begin{equation}\label{phspa} 
  \frac{V^N}{(2\pi)^{3N}} \Bigg\{ \prod_j \frac{1}{N_j!} \Bigg[
  \int \d^3 p \Bigg]^{N_j} \Bigg\} \delta^4(P_i - \sum_i p_i)
\end{equation}
where $N=\sum_j N_j$, and the general expression of the decay rate into the 
channel $\Nj$ of a massive particle (cluster) in relativistic quantum mechanics:
\begin{equation}\label{rate}
  \Gamma_{\Nj} = \frac{1}{(2\pi)^{3N}} \Bigg\{ \prod_j \frac{1}{N_j!} \Bigg[
  \int \frac{\d^3 p}{2\epsilon_j} \Bigg]^{N_j} \Bigg\} \delta^4(P - \sum_i p_i)
  | M_{fi} |^2
\end{equation}  
where $| M_{fi} |^2$ is the Lorentz-invariant dynamical matrix element governing
the decay. Assuming, for sake of simplicity, all spinless particles, $| M_{fi} |^2$ 
may in principle depend on all relativistic invariants 
formed out of the four-momenta of the $N$ particles, as well as on all possible 
isoscalars formed out of the isovector operators ${\bf I}_i$. Suppose that 
$| M_{fi} |^2 = \alpha^N$, so that the whole dynamics reduces to introduce the 
same multiplicative constant $\alpha$ for each particle in the channel. Then, it 
is possible to calculate quite easily the generating function of the multi-particle 
multiplicity distribution starting from Eq.~(\ref{rate}):
\begin{eqnarray}\label{generat}
  G(\lambda_1,\ldots,\lambda_K) &=& \sum_{\Nj} \Gamma_{\Nj} \prod_j \lambda_j^{N_j} 
  \nonumber \\
  &=& \sum_{\Nj} \Bigg\{ \prod_j \frac{\alpha^{N_j}\lambda_j^{N_j}}{(2\pi)^{3N_j} N_j!} 
  \Bigg[\int \frac{\d^3 p}{2\epsilon_j} \Bigg]^{N_j} \Bigg\} \, \delta^4(P - \sum_i p_i)
  \nonumber \\
  &=& \frac{1}{(2\pi)^4} \int \d^4 x \; \e^{\i P \cdot x}
  \exp \left[ \sum_j \frac{\alpha}{(2\pi)^3}\int \frac{\d^3 p}{2\epsilon_j} \; 
  \e^{-\i p_j\cdot x} \lambda_j  \right] \nonumber \\  
  &=& \frac{1}{(2\pi\i)^4} \int \d^4 z \; \e^{P \cdot z}
  \exp \left[ \sum_j \frac{\alpha}{(2\pi)^3}\int \frac{\d^3 p}{2\epsilon_j} \; 
  \e^{- p_j \cdot z} \lambda_j  \right]    
\end{eqnarray}
where a Fourier decomposition of the four-dimensional delta has been used and
the integral Wick rotated by using $z=\i x$. 
If $P^2$ is sufficiently large, we can expand the above integral around the 
saddle-point $z_0$ obtained by solving the equation:
\begin{equation}
  P + \frac{\partial}{\partial z} 
  \sum_j \frac{\alpha}{(2\pi)^3}\int \frac{\d^3 p}{2\epsilon_j} \; \e^{- p_j
  \cdot z} \lambda_j = 0
\end{equation}
If $P=(M,{\bf 0})$ it is not difficult to realize that $z_0 = (\beta,{\bf 0})$
and the generating function can be approximated as:
\begin{equation}
   G(\lambda_1,\ldots,\lambda_K) \sim \exp \left[ \sum_j \frac{\alpha}{(2\pi)^3}\int 
   \frac{\d^3 p}{2\epsilon_j} \e^{- \beta \epsilon_j } \lambda_j \right] 
\end{equation}
so that the mean number of particles of the species $j$ reads:
\begin{equation}\label{quasi}
   \langle n \rangle_j = \frac{\alpha}{(2\pi)^3}\int 
   \frac{\d^3 p}{2\epsilon_j} \; \e^{- \beta \epsilon_j} 
\end{equation}
which is very similar to a thermal distribution:
\begin{equation}\label{proper}
   \langle n \rangle_j = \frac{V}{(2\pi)^3}\int \d^3 p \; \e^{- \beta \epsilon_j} 
\end{equation}
were not for the different measure in the momentum integral. As it should be clear
from its derivation, the constant $\beta$ in Eq.~(\ref{quasi}) is certainly not a 
temperature, rather a soft scale parameter which is related to the effective finite 
interaction range. Yet, the ratios of average multiplicities of particles of different 
species mimic a thermodynamic behaviour. This is the so-called {\em phase space 
dominance}. The authors of ref.~\cite{oregon} work out 
a more specific example based on QED and they conclude, quite reasonably, that a 
fairly good fit to particle multiplicities may be provided if integral expressions 
like (\ref{quasi}) are used instead of an actual Boltzmann integral. I want to even 
reinforce their statement by adding that the
actual fits to particle multiplicities in \ee, pp and other collisions relied on
supplementary assumptions which are not expected to be exact in a basic statistical
model framework, so that deviations from the ``pure" statistical model predictions
may arise which can be of the same order of the difference between the actual SHM
and the formula (\ref{quasi}). 

Thus, I subscribe to the argument in ref.~\cite{oregon} but it should be stressed 
that the just described phase space dominance is a highly non-trivial assumption. 
In fact, the recovery of a thermal-like expression like (\ref{quasi}) ought to a 
very special form of the matrix element $|M_{fi}|^2$, where both the dependence
on kinematical and isospin invariants was disregarded. If a different form, still 
perfectly legitimate and possible, is assumed, the thermal-like behaviour is  
spoiled. For instance, one could have:
\begin{equation}\label{special} 
 |M_{fi}|^2 \propto \alpha^3 M f(\alpha m_1) \cdot \ldots \cdot \alpha^3 M f(\alpha m_N) 
 \cdot g(I_1) \cdot \ldots \cdot g(I_N)
\end{equation}
with a generic factor $f(\alpha m_j)g(I_j)$ for each particle depending on its mass 
$m_j$ and its isospin $I_j$ and on a single scale $\alpha$ whose dimension is the 
inverse of an energy; the factors $\alpha^3 M$ in Eq.~(\ref{special}) are 
introduced in order to make the average particle multiplicities in the large 
multiplicity limit proportional to the mass of the cluster $M$:
\begin{equation}\label{quasi2}
   \langle n \rangle_j = \frac{\alpha^3 M f(m_j)g(I_j)}{(2\pi)^3}\int 
   \frac{\d^3 p}{2\epsilon_j} \e^{- \beta \epsilon_j} 
\end{equation}
It is not difficult to realize that the production function Eq.~(\ref{quasi2}) 
might be dramatically different from the thermal one. It should be emphasized that 
also a factorizable dynamical matrix element depending only on masses and isospins 
like in Eq.~(\ref{special}) is quite an exceptional one. In fact, in principle, 
there could be dependence on other independent invariants like $(p_i + p_j)^2$, ${\bf I}_i
\cdot {\bf I}_j$, ${\bf I}_i \cdot ({\bf I}_j \times {\bf I}_k)$ etc. Therefore, an 
observed phase space dominance in multihadron production is not a trivial fact 
and tells us something important about the characteristics of the underlying non 
perturbative QCD dynamics, besides providing us with an empirically good model. 

Similar arguments are presented in ref.~\cite{rischke} and, more extensively, 
in ref.~\cite{koch} where the concept of phase space dominance is even more 
explicitely defined. There, quantities like the previous $\beta$ arising from a 
saddle-point asymptotic expansion and mimicking a temperature are called ``Lagrange 
multipliers" just to emphasize the difference from an actual temperature. 
However, in the effort of analyzing the meaning of the statistical model results, 
two very questionable statements are introduced:
\begin{itemize}
\item that the so-called Lagrange multipliers have no physical meaning even
for a properly defined phase space integral like (\ref{phspa});
\item that the phase space dominance is trivial when the average multiplicities
are very large.
\end{itemize}
For the second point, the counter-argument is straightforward: just take $f(m)=
A\exp(Cm^2)$ (though odd it might look) or, alternatively, $g(I)= AI^2 + C$ in 
Eq.~(\ref{quasi2}) with $A,C$ positive constants depending on centre-of-mass 
energy and the thermal shape of mass production function is destroyed {\em for 
any multiplicity}.
 
The first point is more subtle and requires a somewhat general discussion because 
there seems to be some confusion as to what deserves to be called ``thermal" 
and, conversely, what is only ``statistical". If the word ``statistical" is used to 
to mean some property of the dynamical matrix element of being independent
of most kinematical variables, like that leading to Eq.~(\ref{quasi}), then of 
course it has nothing to do with a proper thermal thing. If, on the other hand, the 
word ``statistical" means, like in SHM, equal probability in phase space, where phase 
space is appropriately measured with $\d^3 x \d^3 p$ for any particle like in 
Eq.~(\ref{phspa}), and a volume is involved, then ``statistical" and ``thermal" 
can be taken as synonimous (for purists only for sufficiently large volumes)
because there is no quantitative difference between them. In fact, what makes 
the difference between Eq.~(\ref{quasi}) and a proper thermal formula is the 
measure in momentum space and the absence of a volume. If, in a proper 
statistical mechanical framework, the two conditions of statistical equilibrium 
and finite volume are met, temperature can be defined (e.g. through a saddle point
expansion) no matter how the system got to statistical equilibrium and even in
absence of an external bath. Many authors (e.g. \cite{gross}) take the definition
$T^{-1} = \partial S/\partial E$ where $S$ is the entropy, a well defined quantity
for any closed system. All other definitions of temperature, be a Lagrange
multiplier for the maximization of entropy at a fixed energy \cite{heinzlect}, 
or a saddle point of 
the microcanonical partition function, should converge to the same value in the 
limit of large volumes and are therefore physically meaningful temperature. 
Macroscopically inspired definitions requiring physical exchange of energy with 
a heat reservoir are too restrictive, and certainly not suitable for heavy
ion collisions as well, where such a heat reservoir does not exist. On the other
hand, these definitions must coincide with the most general definition based 
on statistical mechanics. 

In the same spirit, some authors \cite{heinzlect} try to make clear a distinction 
between the temperature determined in the SHM by fitting particle abundances and 
a ``proper" temperature which would be achieved through inelastic reinteractions 
of formed particles. The former is called Lagrange multiplier for the maximization 
of entropy, just to emphasize the difference. Again, I would like to stress that 
there is no actual quantitative difference between those two temperatures so that a 
hadronization temperature, if confirmed, can be properly called a temperature. One
can certainly make a distinction as to how statistical equilibrium was achieved,
which is as important as the statistical equilibrium itself, but if energy is equally 
shared among all possible states within a finite (possibly
large) volume, temperature is temperature no matter how the system got to statistical 
equilibrium. What would make the exponential parameter fitted in the framework of 
SHM different from an actual temperature can be only a quantitative difference, 
like e.g. the difference between $\beta$ in Eq.~(\ref{quasi}) and $\beta$ in 
Eq.~(\ref{proper}).

What can be done then to distinguish between a genuine statistical-thermal model and 
other possible pseudo-statistical models like the one leading to the formula 
(\ref{quasi})? Besides kinematical features, it would be desirable to bring out
effects related to the finite volume, which is a peculiarity of the statistical model.
Indeed, the study of average inclusive multiplicities or inclusive 
$p_T$ spectra does not allow clearcut conclusions because those observables are not 
sensitive enough to different integration measures (i.e. $V \d^3 p$ versus 
$\d^3 p/2\epsilon$) and much information is integrated away. A much more effective 
test would be studying the rates of exclusive channels, i.e. 
$\Gamma_{\Nj}/\Gamma_{\{N'_j\}}$, which are much more sensitive to the integration
measure in the momentum integrals and the shape of dynamical matrix element. 
Unfortunately, exclusive channels can be measured only at low energy (some GeV) where 
none of the conservation laws, including angular momentum, parity and isospin, can 
be neglected, as pointed out in ref.~\cite{heinz}
where \ppb annihilation at rest has been studied in this framework. This makes 
calculations rather cumbersome and difficult from the numerical point of view.
None of the numerous previous studies in literature has tackled the problem
without introducing approximations unavoidably implying large errors in the 
calculations. Fully microcanonical calculations including both four-momentum and 
angular momentum conservation have not ever been done, and only recently the increased 
computing power and purposely designed techniques allowed the calculation of
averages in the microcanonical ensemble, yet only with energy and momentum 
conservation \cite{weai,bf2}.   
  
\section{What is the meaning of it?}

Now that we have discussed in some detail the foundations of the statistical
model, and possible interpretations of its success, we are finally left with an 
inevitable question: what is the meaning of this model in the framework of the 
basic theory of strong interactions, QCD? Otherwise stated, is it possible to 
show from a more fundamental theory that extended massive objects such 
as clusters exist and that the statistical filling of their multihadronic phase 
space effectively occurs? Or, alternatively, that QCD implies a similar phenomenon
(though quantitatively distinct and distinguishable), called {\em phase space 
dominance}?
As yet, we are not able to answer this question because QCD has not been solved
in the non-perturbative regime. Therefore, we will try to argue about some simpler 
issue.
 
A first issue is the meaning of the mixture of states (\ref{mix1}) that we 
have used to describe cluster decays. From a quantum mechanical viewpoint, a mixture 
of states is only a mean to describe our ignorance of the state of the system, 
which is always supposed to be a pure one, be it entangled or not. We do not want 
here to slip into fundamental quantum mechanics problems like decoherence and 
measurement, which may render a mixture of states an {\em objective} description 
of the system. Just to make this issue a concrete one in our perspective, it 
suffices to mention a (low energy) collision creating one cluster: of course this
should be described with a pure state.   

Let $\ket{i}$ the pure quantum state of a cluster; we can instance think of this
state as that which can be calculated in the bag model in terms of free parton
fields states confined within a finite region. We can write the transition 
amplitude to a localized multi-hadronic state within the cluster:
\begin{equation}\label{trans1}
  \bra{h_V} T \ket{i} \propto \bra{h_V} T \Pro_i \ket{i} = \bra{\Pro_i h_V} T \ket{i} 
\end{equation}
where the last equality follows from the conservation laws, that is the transition
operator $T$ depends on the hamiltonian of strong interactions and ought to commute 
with the projector onto conserved quantities. We can build up a basis of the Hilbert 
space including the $\ket{h_V}$ vectors by adding to them the multihadronic 
states localized outside $V$, i.e. the region denoted with $\bar{V}$. We can then 
write:
\begin{equation}\label{resol}
  {\sf I} = \sum_{h_V} \ket{h_V}\bra{h_V} + \sum_{h_{\bar V}} 
  \ket{h_{\bar V}}\bra{h_{\bar V}} 
\end{equation}
Essentially, the results of the statistical model can be recovered by assuming:
\begin{eqnarray}\label{statmod}
 && \bra{h_{\bar V}} T \ket{i} = 0  \qquad \forall \ket{h_{\bar V}} \nonumber \\
 && |\bra{\Pro_i h_V} T \ket{i}|^2 \equiv | c_{h_V} |^2 = C
\end{eqnarray}
where $C$ is a constant, independent of the state $\Pro_i \ket{h_V}$. The first
of the two equations in (\ref{statmod}) states that no transition can occur to
a state outside the cluster volume; the second, that the transition probability
is uniform for all localized states with the same quantum numbers as the cluster
itself. In a sense, these statements amount to restate the Hagedorn's hypothesis 
of a resonance as being made of a uniform superposition of hadrons and resonances. 
From the previous assumptions and using Eqs.~(\ref{trans1}),(\ref{resol}), one
can calculate the transition amplitude to an asymptotic state $\ket{f}$:
\begin{equation}
  \bra{f} T \ket{i} = \bra{f} \left( \sum_{h_V} \ket{h_V}\bra{h_V} + 
  \sum_{h_{\bar V}} \ket{h_{\bar V}}\bra{h_{\bar V}}\right) T \ket{i} =  
  \sum_{h_V} \braket{f}{h_V}\bra{\Pro_i h_V} T \ket{i} = 
  \sum_{h_V} \braket{f}{h_V} c_{h_V}
\end{equation}
so that:
\begin{equation}
  |\bra{f} T \ket{i}|^2 = |\sum_{h_V} \braket{f}{\Pro_i h_V} c_{h_V} |^2 = 
  \sum_{h_V} |\braket{f}{\Pro_i h_V}|^2 C + \sum_{h_V \ne h'_V} 
  \braket{f}{\Pro_i h_V}\braket{\Pro_i h'_V}{f} c_{h_V}c*_{h'_V}
\end{equation}
The first term in the right hand side of above equation is just proportional
to (\ref{best}). So, the statistical model results are fully recovered if:
\begin{equation}\label{vanish}
 \sum_{h_V \ne h'_V} \braket{f}{P_i h_V}\braket{P_i h'_V}{f} c_{h_V}c*_{h'_V}
  \simeq 0
\end{equation}
or, in other words, if the amplitudes $c_{h_V} = \sqrt{C} \exp (\i \phi_{h_V})$ 
defined in (\ref{statmod}) have random phases $\phi_{h_V}$, so to make the 
cross-term sum vanishing. 
 
Hence, we have actually rephrased the question whether the statistical model
can be an effective model for the hadronization process actually driven by QCD
on the question whether the conditions~(\ref{statmod}) and (\ref{vanish}) apply
in a QCD-inspired picture.
Since it is presently not possible to answer to this question either, we are left
with the more approachable problem of verifying the predictions of the statistical
model more thouroughly, as we have discussed at the end of previous section.
 
\section{Conclusions}
 
We have discussed in some detail the ideas and the interpretations of the 
success of the statistical model in reproducing soft observables in high energy
collisions. It is certainly crucial to understand the why of this success from firts
QCD principles, but in the meantime it is useful to stick to a more pragmatic
attitude and ask ourselves whether we can test this model more deeply than 
what has been done as yet. Particularly, by testing the model against exclusive
channel rates, we can assess whether the thermal-like features of inclusive 
particle production show up at high energy because of the quasi-independence of 
dynamical matrix elements in the soft non-perturbative regime ({\em phase space 
dominance}). In fact, it is difficult to bring out deviations from a genuine 
statistical model from the analysis of inclusive quantities only because too
much information is integrated away. On the other hand, such deviations should 
show up in more detailed observables, like, e.g., exclusive channel rates. 
In this regard, relevant data are available only at low energy (some GeV in 
centre-of-mass frame) and this requires the implementation of full microcanonical 
calculations, which have never been done without introducing too drastic 
approximations. We have outlined an appropriate framework (in Sect.~2) for the 
full microcanonical formulation of the model, on the basis of group projection 
techniques. This the first step to implement the calculation; numerical 
work is currently ongoing.

\section*{Acknowledgements}

The author warmly thanks the organizers of the workshop {\it Focus on Multiplicity}
for having provided a stimulating environment. Part of this work originated from 
discussions with V. Koch. The author gratefully acknowledges T. Gabbriellini 
for stimulating and clarifying discussions. 
 
\section*{References}

\smallskip
  
\end{document}